\documentclass[12pt]{iopart}
\usepackage{graphicx}
\usepackage{iopams}

\begin{document}

\title[Semigroup techniques in optical quantum information]{Semigroup
  techniques for the efficient classical simulation of optical quantum
  information} 

\author{S D Bartlett}

\address{Department of Physics and Centre for Advanced Computing --
  Algorithms and Cryptography, Macquarie University, Sydney, NSW 2109,
  Australia}

\begin{abstract}
  A framework to describe a broad class of physical operations
  (including unitary transformations, dissipation, noise, and
  measurement) in a quantum optics experiment is given.  This
  framework provides a powerful tool for assessing the capabilities
  and limitations of performing quantum information processing tasks
  using current experimental techniques.  The Gottesman-Knill theorem
  is generalized to the infinite-dimensional representations of the
  group stabilizer formalism and further generalized to include
  non-invertable semigroup transformations, providing a theorem for
  the efficient classical simulation of operations within this
  framework. As a result, we place powerful constraints on obtaining
  computational speedups using current techniques in quantum optics.
\end{abstract}

\section{Introduction}

Information processing using the rules of quantum mechanics may allow
tasks that cannot be performed using classical laws~\cite{Nie00}.  The
efficient factorization algorithm of Shor~\cite{Sho94} and secure
quantum cryptography~\cite{BB84} are two examples.  Of the many
possible realizations of quantum information processes, optical
realizations have the advantange of negligible decoherence: light does
not interact with itself, and thus a quantum state of light can be
protected from becoming entangled with the environment.  Several
proposed optical schemes~\cite{Chu95,KLM01,Got01b,Llo99} offer
significant potential for quantum information processing.

In order to prove theorems regarding the possibilities and limitations
of optical quantum computation, one must construct a framework for
describing all types of physical processes (unitary transformations,
projective measurements, interaction with a reservoir, etc.)\ that can
be used by an experimentalist to perform quantum information
processing.  Most frameworks currently employed (e.g.,~\cite{Llo99})
are restricted to describing only unitary transformations.  However,
such transformations are a subset of all possible physical processes.
Non-unitary transformations such as dissipation, noise, and
measurement must also be described within a complete framework.  The
new results of Knill et al~\cite{KLM01} show that photon counting
measurements allow for operations that are ``difficult'' with unitary
transformations alone; thus, non-unitary processes may be a powerful
resource in quantum information processing and must be considered in
any framework that attempts to address the capabilities of quantum
computation with optics.

In this paper, we show that unitary transformations, measurements and
any other physical process can be described in the unified formalism
of completely positive (CP) maps.  Also, a broad class of these maps
which includes linear optics and squeezing transformations, noise
processes, amplifiers, and measurements with feedforward that are
typical to quantum optics experiments can be described within the
framework of a \emph{Gaussian semigroup}.  This framework allows us to
place limitations on the potential power of certain quantum
information processing tasks.

One important goal is to identify classes of processes that can be
efficiently simulated on a classical computer; such processes cannot
possibly be used to provide any form of ``quantum speedup''.  The
Gottesman-Knill (GK) theorem~\cite{Got99,Nie00} for qubits and the CV
classical simulatability theorems of Bartlett et
al~\cite{Bar02b,Bar02c} provide valuable tools for assessing the
classical complexity of a quantum optical process.  It is shown here
that semigroup techniques provide a powerful formalism with which one
can address issues of classical simulatability.  In particular, a
classical simulatability result is presented for a general class of
quantum optical operations, and thus a no-go theorem for quantum
computation with optics is proven using semigroup techniques.

\section{Semigroup Description of Gaussian operations}

Consider an optical quantum information process involving $n$ coupled
electromagnetic field modes, with each mode described as a quantum
harmonic oscillator.  The two observables for the (complex) amplitudes
of a single field mode serve as canonical operators for this
oscillator.  A system of $n$ coupled oscillators, then, carries an
irreducible representation of the Heisenberg-Weyl algebra hw($n$),
spanned by the $2n$ canonical operators $\{ q_i,p_i,i=1,\ldots,n\}$
along with the identity operator $I$.  These operators satisfy the
commutation relations $[q_i,p_j] = i\hbar \delta_{ij} I$.  We express
the $2n$ canonical operators in the form of a phase space vector
$\bi{z}$ with components $z_i = q_i$ and $z_{n+i} = p_i$ for
$i=1,\ldots,n$.  These operators satisfy $[z_i,z_j] =
i\hbar\Sigma_{ij}$, with $\Sigma$ the skew-symmetric $2n \times 2n$
matrix
\begin{equation}
  \label{eq:SigmaMatrix}
  \Sigma = \left ( \begin{array}{cc}
             0 & I_n \\ 
             -I_n & 0 \\ 
           \end{array} \right )
\end{equation}
and $I_n$ the $n \times n$ identity matrix.  For a state $\rho$
represented as a density matrix, the \emph{means} of the canonical
operators is a vector defined as the expectation values
$\boldsymbol{\xi} = \langle\bi{z}\rangle_\rho$, and the
\emph{covariance matrix} is defined as
\begin{equation}
  \label{eq:CovarianceMatrix}
  \Gamma = \langle (\bi{z} -
  \boldsymbol{\xi})(\bi{z} -\boldsymbol{\xi})^\dag \rangle_\rho
  - i \Sigma \, .
\end{equation}
A \emph{Gaussian state} (a state whose Wigner function is Gaussian and
thus possesses a quasiclassical description) is completely
characterized by its means and covariance matrix~\cite{Lin00}.
Coherent states, squeezed states, and position- and
momentum-eigenstates are all examples of Gaussian states.

We define $\mathcal{C}_n$ to be the group of linear transformations of
the canonical operators $\{z_i \}$~\cite{Bar02b}; this group
corresponds to the infinite-dimensional (oscillator) representation of
the ``Clifford group'' employed by Gottesman~\cite{Got99}.  For a
system of $n$ oscillators, it is the unitary representation of the
group ISp($2n,\mathbb{R}$) (the inhomogeneous linear symplectic group
in $2n$ phase space coordinates)~\cite{Wun02} which is the semi-direct
product of phase-space translations (the Heisenberg-Weyl group
HW($n$)) plus one- and two-mode squeezing (the linear symplectic group
Sp($2n,\mathbb{R}$)).  Phase space displacements are generated by
Hamiltonians that are linear in the canonical operators; a
displacement operator $X(\boldsymbol{\alpha}) \in$ HW($n$) is defined
by a real $2n$-vector $\boldsymbol{\alpha}$. A symplectic
transformation $M(A) \in$ Sp($2n,\mathbb{R}$), with $A$ a real matrix
satisfying $A^\dag \Sigma A = \Sigma$, is generated by a Hamiltonian
that is a homogeneous quadratic polynomial in the canonical operators.
A general element $C \in \mathcal{C}_n$ can be expressed as a product
$C(\boldsymbol{\alpha},A) = X(\boldsymbol{\alpha})M(A)$, and
transforms the canonical operators as
\begin{equation}
  \label{eq:Clifford}
  C(\boldsymbol{\alpha},A) : \bi{z} \to \bi{z}' = \bi{z}A +
  \boldsymbol{\alpha} \, .
\end{equation}

The group $\mathcal{C}_n$ consists of unitary transformations that map
Gaussian states to Gaussian states; however, unitary transformations
do not describe all physical processes.  In the following, we include
other (non-unitary) CP maps that correspond to processes such as
dissipation or measurement.  We define the \emph{Gaussian semigroup},
denoted $\mathcal{K}_n$, to be the set of Gaussian CP
maps~\cite{Lin00} on $n$ modes: a Gaussian CP map takes any Gaussian
state to a Gaussian state.  Because Gaussian CP maps are closed under
composition but are not necessarily invertible, they form a semigroup.
A general element $T \in \mathcal{K}_n$ is defined by its action on
the canonical operators as
\begin{equation}
  \label{eq:ActionOfSemigroupOnCanonical}
  T(\boldsymbol{\alpha},A,G): \bi{z} \to \bi{z}' =  \bi{z}A +
  \boldsymbol{\alpha} + \boldsymbol{\eta} \, ,
\end{equation}
where $\boldsymbol{\alpha}$ is a real $2n$-vector, $A$ and $G$ are
$2n\times 2n$ real matrices, and $A$ is no longer required to be
symplectic.  Eq.~(\ref{eq:ActionOfSemigroupOnCanonical}) includes the
transformations (\ref{eq:Clifford}) plus additive noise
processes~\cite{Gar00} described by quantum stochastic noise operators
(the vector $\boldsymbol{\eta}$) with expectation values equal to
zero and covariance matrix
\begin{equation}
  \label{eq:NoiseCovariance}
  \langle \boldsymbol{\eta} \boldsymbol{\eta}^\dag \rangle_{\rho_R} 
  - i \Sigma = G - i A^\dag \Sigma A \, .
\end{equation}
Here, $\rho_R$ is a Gaussian `reservoir' state which, in
order to define a CP map, must be chosen such that the noise operators
satisfy the quantum uncertainty relations.  This condition is
satisfied if the noise operators define a positive definite density
matrix, which leads to the condition
\begin{equation}
  \label{eq:CPcondition}
  G +i \Sigma - i A^\dag \Sigma A \geq 0 \, .
\end{equation}
The group $\mathcal{C}_n$ is recovered for $G=0$.

The action of the Gaussian semigroup on the means and covariance
matrix is straightforward and given by
\begin{equation}
  \label{eq:ActionOfSemigroupOnMeansAndCM}
  T(\alpha,A,G): \cases{\boldsymbol{\xi} \to \boldsymbol{\xi}' =
  \boldsymbol{\xi}A  + \boldsymbol{\alpha}  \\ 
  \Gamma \to \Gamma' = A^\dag \Gamma A + G\, . \\ } 
\end{equation}
Because the means and covariance matrix completely define a Gaussian
state, the resulting action of the Gaussian semigroup on Gaussian
states can be easily calculated via this action.

The Gaussian semigroup $\mathcal{K}_n$ represents a broad framework to
describe several important types of processes in a quantum optical
circuit.  The group $\mathcal{C}_n \subset \mathcal{K}_n$ comprises
the unitary transformations describing phase--space displacements and
squeezing (both one-- and two--mode).  Introduction of noise to the
circuit (e.g., via linear amplification) is also in $\mathcal{K}_n$.
Furthermore, the Gaussian semigroup describes certain measurements in
the quantum circuit.  These include measurements where the outcome is
discarded (thus evolving the system to a mixed state) or retained
(where the system follows a specific quantum trajectory defined by the
measurement record~\cite{Car93}).  Finally, the Gaussian semigroup
includes Gaussian CP maps conditioned on the outcome of such
measurements.  For details and examples of all of these types of
Gaussian semigroup transformations, see~\cite{Bar02c}.

\section{Classical Simulation of Gaussian Semigroup Processes}

Using the framework of the Gaussian semigroup, it is straightforward
to prove the classical simulatability result of Bartlett and
Sanders~\cite{Bar02c}.

\medskip
\noindent \textit{Theorem:} Any quantum information process that
initiates in a Gaussian state and that performs only Gaussian
semigroup maps can be \emph{efficiently} simulated using a classical
computer.

\medskip
\noindent \textit{Proof:} Recall that any Gaussian state is completely
characterized by its means and covariance matrix.  For any quantum
information process that initiates in a Gaussian state and involves
only Gaussian semigroup maps, one can follow the evolution of the
means and the covariance matrix rather than the quantum state itself.
For a system of $n$ coupled oscillators, there are $2n$ independent
means and $2n^2+n$ elements in the (symmetric) covariance matrix;
thus, following the evolution of these values requires resources that
are polynomial in the number of coupled systems. \hfill \textit{QED}
\medskip

Because most current experimental techniques in quantum optics are
describable by Gaussian semigroup maps, this theorem places a powerful
constraint on the capability of achieving quantum computational
speedups (tasks that are not efficient on any classical machine) using
quantum optics.


\section{Conclusions}

Semigroup techniques provide a powerful tool for constructing and
assessing new quantum information protocols using quantum optics.
These techniques have been used to show that algorithms or circuits
consisting of only Gaussian semigroup maps can be efficiently
simulated on a classical computer, and thus do not provide the ability
to perform quantum information processing tasks efficiently that
cannot be performed efficiently on a classical machine.  Eisert et
al~\cite{Eis02} use related techniques to show that local Gaussian
semigroup transformations are insufficient for distilling
entanglement: an important process for quantum communication and
distributed quantum computing.  Most current quantum optics
experiments consist only of Gaussian semigroup transformations; thus,
the challenge is to exploit this semigroup to prove new theorems,
limitations and possibilities for quantum information processing using
optics.

\ack 
This project has been supported by Macquarie University and the
Australian Research Council.  The author thanks B.\ C.\ Sanders for
helpful discussions.

\end{document}